\documentclass{elsart}
\usepackage{amsmath,amssymb}
\usepackage{times}
\usepackage{graphicx}
\newtheorem{theorem}{Theorem}
\newtheorem{definition}[theorem]{Definition}

\newtheorem{example}[theorem]{Example}
\newtheorem{remark}[theorem]{Remark}

\def\-{\overline}

\begin{document}

\begin{frontmatter}


\title{Computation of vector sublattices and minimal lattice-subspaces of $\mathbb{R}^k$. Applications in finance.}

\author[label1]{V.N. Katsikis\corauthref{cor1}}
\corauth[cor1]{Vasilios N. Katsikis} \ead{vaskats@gmail.com}
\author[label2]{I.A. Polyrakis}
\ead{ypoly@math.ntua.gr}

\address[label1]{General Department of Mathematics,
Technological Education Institute of Piraeus, 12244, Athens, Greece}
\address[label2]{Department of Mathematics,
   National Technical University of
Athens, Zographou 15780, Athens, Greece}

\begin{abstract}
In this article we perform a computational study of Polyrakis
algorithms presented in \cite{POLY8,POLY9}. These algorithms are
used  for the determination of the vector sublattice and the minimal
lattice-subspace generated by a finite set of positive vectors of
$\mathbb{R}^k$. The study demonstrates that our findings can be very
useful in the field of Economics, especially in completion by
options of security markets and portfolio insurance.
\end{abstract}
\end{frontmatter}

\section{Introduction}\label{S0}
This paper provides computational methods for the determination of
the vector sublattice and the minimal lattice-subspace of
$\mathbb{R}^k$ generated by a set $B=\{x_1,x_2,...,x_n\}$ of
linearly independent positive vectors of $\mathbb{R}^k$. In order to
reach our goal the study of a vector-valued function $\beta$ is
further involved, which is defined and studied by I. Polyrakis in
\cite{POLY8,POLY9}. In particular, in \cite{POLY8,POLY9}, it is
proved that if $B=\{x_1,x_2,...,x_n\}$ is a set of linearly
independent positive vectors of $\mathbb{R}^k$ and
$\beta(i)=\frac{r(i)}{\|r(i)\|_1}$  for each $i=1,2,...,k$, where
$r(i)=(x_1(i),x_2(i),...,x_n(i))$, $R(\beta)$ denotes the range of
the function $\beta$ and $K$ is the convex hull of $R(\beta)$ then
it holds,
\begin{enumerate}
\item[(i)] There exists an $n-$dimensional minimal lattice-subspace
of $\mathbb{R}^k$ containing $B$,
\item[(ii)] The vector sublattice generated by $B$ is an $m-$dimensional
subspace of $\mathbb{R}^k$.
\end{enumerate}
where $n$ is the number of vertices of $K$ and $m$ is the number of
different values of $\beta $.

In \cite{KOUN}, the vector sublattice generated by a set of linearly
independent positive vectors of $\mathbb{R}^k$ was used in order to
provide an answer to the challenging question of market completion
by options, in a discrete-time framework. Also, in \cite{POLY1} it
is proved that if for a set of linearly independent positive vectors
of $\mathbb{R}^k$ there exists a minimal lattice-subspace containing
this set, then one can use the produced subspace for the study of a
cost-minimization problem that is also known as the minimum cost
portfolio insurance problem. From the computational point of view,
several methods are presented in \cite{KATS,KATS1,KATS2} in order to
check whether a finite collection of linearly independent positive
vectors(resp. functions) of $\mathbb{R}^k$(resp. $C[a,b]$) forms a
lattice-subspace or a vector sublattice. In addition, these methods
have been used for the calculation of the minimum-cost portfolio
insurance. The main goals of this article are outlined as
follows:\begin{enumerate}
\item Exploit the theory of positive bases and derive
new computational methods for the calculation of the completion by
options of a two-period security market,
\item Calculate the minimal lattice-subspace generated by a finite
collection of linearly independent positive vectors.
\end{enumerate}
The proposed numerical solution, based upon previous works of the
authors, provide the exact solution to the problem of calculating
the market completion by options and the minimum-cost insured
portfolio. Also, a closer look at Example \ref{Ex5} can tell us that
a manual procedure in order to determine the completion of the given
security market can easily be a prohibited task. In the economics
models that we are working the input data, as for example the
payoffs of the  different securities, are standard for a long time
period and so, by our methods we determine the exact solution for
these problems.

In the present work, the numerical tasks have been performed using
the high-end Matlab programming language. Specifically, the Matlab
7.4 (R2007b) environment \cite{MATL1,MATL2} was used on an Intel(R)
Pentium(R) Dual CPU T2310 @ 1.46 GHz 1.47 GHz 32-bit system with 2
GB of RAM memory running on the Windows Vista Home Premium Operating
System.

\section{Lattice-subspaces and vector sublattices of $\mathbb{R}^k$}\label{S1}
\subsection{Preliminaries}\label{S11}
We first recall some definitions and notation from the vector
lattice theory. Let $\mathbb{R}^k=\{x=(x(1),x(2),...,x(k))|x(i)\in
\mathbb{R}, \textit{ for each } i\},$ where we view $\mathbb{R}^k$
as an ordered space. The \textit{ pointwise order } relation in
$\mathbb{R}^k$ is defined by $$x\leq y \textit{ if and only if }
x(i)\leq y(i) , \textit { for each } i=1,...,k.$$  The positive cone
of $\mathbb{R}^k$ is defined by $\mathbb{R}^k_+=\{x\in
\mathbb{R}^k|x(i)\geq 0, \textit{ for each } i\}$ and if we suppose
that $X$ is a vector subspace of $\mathbb{R}^k$ then $X$ ordered by
the pointwise ordering is an \textit{ordered subspace} of
$\mathbb{R}^k$ with positive cone $X_+=X\cap \mathbb{R}^k_+.$ A
point $x\in \mathbb{R}^k$ is an \textit{ upper bound (resp. lower
bound)} of a subset $S\subseteq \mathbb{R}^k$ if and only if $y\leq
x(\textnormal{resp. }x\leq y)$, for all $y\in S.$ For a two-point
set $S=\{x,y\},$ we denote by $x\vee y (\textnormal{resp. } x\wedge
y)$ the \textit{ supremum } of $S$ i.e., its least upper bound(resp.
the \textit{ infimum } of $S$ i.e., its greatest lower bound). Thus,
$x\vee y (\textnormal{resp. }x\wedge y)$ is the componentwise
maximum(resp. minimum) of $x$ and $y$ defined by
$$(x\vee y)(i)=\max\{x(i),y(i)\}( (x\wedge y)(i)=\min\{x(i),y(i)\}
), \: \textit{ for all } i=1,...,k.$$ An ordered subspace $X$ of
$\mathbb{R}^k$ is a \textit{ lattice-subspace } of $\mathbb{R}^k$ if
it is a vector lattice in the induced ordering, i.e., for any two
vectors $x,y\in X$  the supremum and the infimum of $\{x,y\}$ both
exist in $X.$ Note that the supremum and the infimum of the set
$\{x,y\}$ are, in general, different in the subspace from the
supremum and the infimum of this set in the initial space. An
ordered subspace $Z$ of $\mathbb{R}^k$ is a \textit{ vector
sublattice } or a \textit{ Riesz subspace } of $\mathbb{R}^k$ if for
any $x,y\in Z$ the supremum and the infimum of the set $\{x,y\}$ in
$\mathbb{R}^k$ belong to $Z.$

Assume that $X$ is an ordered subspace of $\mathbb{R}^k$ and
$B=\{b_1,b_2,...,b_n\}$ is a basis for $X.$ Then $B$ is a \textit{
positive basis } of $X$ if for each $x\in X$ it holds that $x$  is
positive if and only if its coefficients in the basis $B$ are
positive. In other words, $B$ is a positive basis of $X$ if the
positive cone $X_+$ of $X$ has the form,
$$X_+=\{x=\sum _{i=1}^n \lambda _ib_i|\lambda _i\geq 0, \textit{ for
each } i\}.$$ Then, for any $x=\sum _{i=1}^n\lambda _ib_i$ and
$y=\sum _{i=1}^n \mu _i b_i$ we have $x\leq y$ if and only if
$\lambda _i\leq \mu _i$ for each $i=1,2,...,n.$

Each element $b_i$ of the positive basis of $X$ is an extremal point
of $X_+$ thus a positive basis of $X$ is unique in the sense of
positive multiples. Recall that a nonzero element $x_0$ of $X_+$ is
an \textit{ extremal point } of $X_+$ if, for any $x\in X, 0\leq
x\leq x_0$ implies $x=\lambda x_0$ for a real number $\lambda $. The
existence of positive bases is not always ensured, but in the case
where $X$ is a vector sublattice of $\mathbb{R}^k$ then $X$ has
always a positive basis. Moreover, it holds that an ordered subspace
of $\mathbb{R}^k$ has a positive basis if and only if it is a
lattice-subspace of $\mathbb{R}^k.$ If $B=\{b_1,b_2,...,b_n\}$ is a
positive basis for a lattice-subspace (or a vector sublattice) $X$
then the lattice operations in $X$, namely $x\nabla y$ for the
supremum and $x \vartriangle y$ for the infimum of the set $\{x,y\}$
in $X$, are given by
$$x\nabla y=\sum_{i=1}^n\max\{\lambda_i,\mu_i\}b_i \textit{
and  }x\vartriangle y=\sum_{i=1}^n\min\{\lambda_i,\mu_i\}b_i,$$ for
each $x=\sum _{i=1}^n\lambda _ib_i,  y=\sum _{i=1}^n \mu _i b_i \in
X.$ A vector sublattice is always a lattice-subspace, but the
converse is not true as shown in the next example,

\begin{example}\label{Ex1}
{\em Let $X=[x_1,x_2,x_3]$ be the subspace of $\mathbb{R}^4$
generated
 by the vectors $x_1=(6,0,0,1),x_2=(6,4,0,0), x_3=(8,4,2,0).$ An
 easy argument shows that the set $B=\{b_1,b_2,b_3\}$ where
 \[\left[\begin{array}{c}
                                                                b_1 \\
                                                                b_2 \\
                                                                b_3 \\
                                                              \end{array}
                                                            \right]=\left[
                                                                      \begin{array}{cccc}
                                                                        2&0&2&0\\
                                                                        12&8&0&0\\
                                                                        6&0&0&1\\
                                                                      \end{array}
                                                                    \right]\]
forms a positive basis of $X$ therefore $X$ is a lattice-subspace of
 $\mathbb{R}^4.$ On the other hand, let us consider the vectors
 $y_1=2x_1+x_2=(18,4,0,2)$ and $y_2=x_3-x_2=(2,0,2,0)$
of $X.$ Then,  $y_1\vee y_2=(18,4,2,2)$ and since
$y_1=\frac{1}{2}b_2+2b_3, y_2=b_1,$ it follows that $y_1\nabla
y_2=b_1+\frac{1}{2}b_2+2b_3=(20,4,4,2)$. From the definition of a
vector sublattice we have that $X$ is a vector sublattice of
$\mathbb{R}^4$ if for each $x,y \in X$ it holds
 $x\vee y=x\nabla y \in X$ and $x\wedge y=x\vartriangle y \in X$.
Therefore,  since there are two elements $y_1,y_2 \in X$ such that
$y_1 \vee y_2\neq y_1\nabla y_2$ it follows that $X$ is not a vector
sublattice of $\mathbb{R}^4.$}
\end{example}

For an extensive presentation of lattice-subspaces, positive bases
and vector sublattices, we refer to \cite{ABRA2,POLY1,POLY8,POLY9},
for computational methods in positive bases theory we refer to
\cite{KATS,KATS1,KATS2} and for several applications in the theory
of finance the reader may refer to \cite{ALIP4,ALIP2,ALIP3,KOUN}.

\subsection{The mathematical problem}\label{S12} In this section we shall present the results of \cite{POLY8,POLY9} for the construction
of a minimal lattice-subspace of $\mathbb{R}^k$ containing a
linearly independent subset of $\mathbb{R}^k_+$ together with the
construction of the vector sublattice of $\mathbb{R}^k$ that this
subset generates. In particular, given a collection of linearly
independent, positive vectors,  $ x_1,x_2,...,x_n $ of
$\mathbb{R}^k$ then the basic tool for our study is a special
function, named basic function, that this collection of vectors
defines. This function was first introduced and studied in
\cite{POLY8}. In our analysis we shall use the notation introduced
in \cite{POLY8}, so let us denote by  $r$  the function
$r:\{1,2,...,k\}\rightarrow \mathbb{R}^k$ such that
$$r(i)=(x_1(i),x_2(i),...,x_n(i))$$
 and by  $\beta $ the function
$\beta:\{1,2,...,k\}\rightarrow \mathbb{R}^k$ such that
$$\beta (i)=\frac{r(i)}{\|r(i)\|_1}$$
for each $i\in \{1,2,...,k\}$ with $\|r(i)\|_1\neq 0.$ We shall
refer to $\beta $ as the \textit{basic function} of the vectors
$x_1,x_2,...,x_n.$ The set \[R(\beta)=\{\beta (i)|i=1,2,...,k,
\textit{ with } \|r(i)\|_1\neq 0\},\] is the \textit{range} of the
basic function and the \textit{cardinal number}, $cardR(\beta),$ of
$R(\beta)$ is the number of different elements of $R(\beta ).$ Let
$cardR(\beta)=m$ then $n\leq m\leq k$ and by $K$ we shall denote the
convex hull of $R(\beta)$ which is, as the convex hull of a finite
subset of $\mathbb{R}^k$, a polytope with $d$ vertices and each
vertex of $K$ belongs to $R(\beta ).$ It is clear that $n\leq d\leq
m.$

An essential issue for our analysis are the conditions under which a
collection of linearly independent, positive vectors
$x_1,x_2,...,x_n$ of $\mathbb{R}^k$ can be used to derive a minimal
lattice-subspace and a vector sublattice containing these vectors.
In \cite{POLY9}, theorem 3.19 is a criterion for lattice-subspaces
and vector sublattices and provides a full answer on the topic on
the basis of describing the geometry of this problem. In the case
where $X=[x_1,x_2,...,x_n]$ is a lattice-subspace or a vector
sublattice of $\mathbb{R}^k$ then the theorem determines a positive
basis in $X$ while, in the opposite case, theorem 3.19 provides a
minimal lattice-subspace and a vector sublattice containing $X$. In
order to state theorem 3.19, let us consider
$R(\beta)=\{P_1,P_2,...,P_m\}$ such that the first $n$ vertices
$P_1,P_2,...,P_n$ are linearly independent and $P_1,P_2,...,P_d$ are
the vertices of $K$, $n\leq d\leq m$. Also, $A^T$ denotes the
transpose matrix of a matrix $A.$

\begin{theorem}\cite[\textit{Theorem 3.19}]{POLY9}.\label{T3} Suppose that the above assumptions are
satisfied. Then,
\begin{enumerate}
\item[$(i)$] $X$ is a vector sublattice of $\mathbb{R}^k$ if and
only if $R(\beta)$ has exactly $n$ points (i.e., $m=n$). Then a
positive basis $b_1,b_2,...,b_n$ for $X$ is defined by the formula
$$ (b_1,b_2,...,b_n)^T =A^{-1}(x_1,x_2,...,x_n)^T,$$ where  $A$ is the
$n\times n$ matrix whose $i$th column is the vector $P_i,$ for
each $i=1,2,...,m.$

\item[$(ii)$] $X$ is a lattice-subspace of
$\mathbb{R}^k$ if and only if the polytope $K$ has $n$ vertices
(i.e., $d=n$). Then a positive basis $b_1,b_2,\ldots ,b_n$ for $X$
is defined by the formula
$$ (b_1,b_2,\ldots ,b_n)^T =A^{-1}(x_1,x_2,\ldots ,x_n)^T,$$ where  $A$ is the
$n\times n$ matrix whose $i$th column is the vector $P_i,$ for each
$i=1,2,\ldots ,d.$

\item[$(iii)$] Let $m>n$. If $I_s=\beta ^{-1}(P_s)$, and \[x_s=\sum_{i\in I_s} \|r(i)\|_1e_i,\:\:s=n+1,n+2,\ldots,m,\]
then \[Z=[x_1,x_2,\ldots ,x_n,x_{n+1},\ldots ,x_m]\] is the vector
sublattice generated by $x_1,x_2,\ldots ,x_n$ and $\dim Z=m.$

\item[$(iv)$] Let $d>n$. If $\xi_i:D(\beta)\rightarrow \mathbb{R}_+,
i=1,2,\ldots ,d$ such that $\sum _{i=1}^d\xi_i(j)=1$ and
$\beta(j)=\sum _{i=1}^d\xi_i(j)P_i$ for each $j\in D(\beta),$ and
$x_{n+i},i=1,2,\ldots ,d-n,$ are the following vectors of
$\mathbb{R}^k:$ \[x_{n+i}=\sum_{j\in
D(\beta)}\xi_{n+i}(j)\|r(j)\|_1e_j,\] then \[Y=[x_1,\ldots
,x_n,x_{n+1},\ldots ,x_d]\] is a minimal lattice-subspace of
$\mathbb{R}^k$ containing $x_1,x_2,\ldots ,x_n$ and $\dim Y=d.$
\end{enumerate}
\end{theorem}

In \cite{KATS2}, a new numerical package is provided so that, for a
variety of dimensions and subspaces, we are able to check conditions
$(i),(ii)$ of Theorem \ref{T3}. In addition, in the case where
$X=[x_1,x_2, \ldots,x_n]$ is a vector sublattice or a
lattice-subspace the proposed method, in \cite{KATS2}, provides the
positive basis of $X$ therefore a complete description of the vector
sublattice or the lattice-subspace is given.

In the case where $X$ is not a vector sublattice or a
lattice-subspace, an essential issue for our analysis is the
construction of a powerful and efficient package in order to
calculate the vector sublattice and a minimal lattice-subspace
containing $X$ by using conditions $(iii),(iv)$ of Theorem \ref{T3}.

\subsection{The algorithm}\label{S13}
\begin{enumerate}
\item[$(1)$] Determine the function $\beta$ as well as the range,
$R(\beta)$, of $\beta.$
\item[$(2)$] Compute the number $m=card R(\beta),$ and  the number $d$ of vertices  of the polytope $K.$
\item[$(3)$] If $n=m$ (vector sublattice case) or $n=d$ (lattice-subspace case) then,  determine a positive basis of $X.$
\item[$(4)$] If $m>n$, then $Z=[x_1,x_2,\ldots ,x_n,x_{n+1},\ldots ,x_m]$ is the vector
sublattice generated by $x_1,x_2,\ldots ,x_n$, where $x_s=\sum_{i\in
I_s} \|r(i)\|_1e_i$ and $I_s=\beta ^{-1}(P_s)$ for each
$s=n+1,n+2,\ldots,m$ .
\item[$(5)$] If $d>n$, then $Y=[x_1,\ldots
,x_n,x_{n+1},\ldots ,x_d]$ is a minimal lattice-subspace of
$\mathbb{R}^k$ containing $x_1,x_2,\ldots ,x_n$, where the vectors
$x_{n+i}=\sum_{j\in D(\beta)}\xi_{n+i}(j)\|r(j)\|_1e_j,$ $i=1,\ldots
, d-n$, were defined in $(iv)$ of Theorem \ref{T3}.
\end{enumerate}

Note that, the steps of this algorithm are based upon Theorem
\ref{T3}. In the following section, we present the translation
followed by the implementation of this algorithm in $\mathbb{R}^k$
within two Matlab-based functions named
 \verb SUBlat  and  \verb MINlat  . These functions, as we shall see in Subsection \ref{S21}, provides
an important tool in order to investigate minimal lattice-subspaces
and vector sublattices of $\mathbb{R}^k$ generated by a set of
linearly independent, positive vectors $x_1,x_2,...,x_n$ of
$\mathbb{R}^k$.

\section{ The computational method}\label{S2}
\subsection{Method presentation and examples}\label{S21}
Our proposed numerical method is based on the introduction of two
functions, namely \verb SUBlat  and  \verb MINlat  (see Appendix),
that enable us to perform fast testing for a variety of dimensions
and subspaces. Also, both of these functions are using the function
\verb SUBlatSUB  from \cite{KATS2}, in order to calculate a positive
basis. For the sake of completeness, we provide the
     \verb SUBlatSUB  function in the Appendix. Recall that, the numbers
$n,m,d,k$ denote the dimension of $X$, the cardinality of
$R(\beta)$, the number of vertices of the convex hull of $R(\beta)$
and the dimension of the initial Euclidean space, respectively.

The function \verb SUBlat  first determines $R(\beta)$ and the
number $m=card R(\beta)$ and then the  $n$ linearly independent
vertices $P_i,\:i=1,\ldots ,n$ of the polytope $K$. Finally, the
program calculates the subsets $I_s=\beta ^{-1}(P_s)$ and then the
vectors
\[x_s=\sum_{i\in I_s} \|r(i)\|_1e_i,\] for each $s=n+1,n+2,\ldots,m.$ Therefore, since by theorem \ref{T3} $(iii)$ the subspace
\[Z=[x_1,x_2,\ldots ,x_n,x_{n+1},\ldots ,x_m]\] is the vector
sublattice generated by the given vectors $x_1,x_2,\ldots ,x_n$ the
program responds, by using the function \verb SUBlatSUB  of
\cite{KATS2}, with two $m\times k$ matrices. The rows of the first
matrix are the vectors   \[x_1,x_2,\ldots ,x_n,x_{n+1},\ldots
,x_m,\] while the rows of the second matrix are the vectors
$b_1,\ldots ,b_m$ of the positive basis for the subspace $Z$.

In order to determine a minimal lattice-subspace generated by the
given collection of vectors we use the function \verb MINlat  . The
\verb MINlat  function first determines $R(\beta)$ and then the
vertices of the polytope $K.$ The correct performance of the
 \verb MINlat    function requires the use of the {\bf convhulln}
Matlab function which is based on Qhull (For information about Qhull
see http://www.qhull.org/). In order to determine a minimal
lattice-subspace of $\mathbb{R}^k$ containing $x_1,x_2,\ldots ,x_n$
the program calculates the vectors $\xi_i:D(\beta)\rightarrow
\mathbb{R}_+, i=1,2,\ldots ,d$ by solving $k$ underdetermined
$n\times d$ linear systems subject to the inequalities $\xi_i\geq 0$
,$i=1,\ldots n.$ Finally, the program defines the vectors,
\[x_{n+i}=\sum_{j\in D(\beta)}\xi_{n+i}(j)\|r(j)\|_1e_j,\:\: i=1,2,\ldots ,d-n,\] and  since the subspace
\[Y=[x_1,\ldots ,x_n,x_{n+1},\ldots ,x_d]\] is a minimal
lattice-subspace containing $x_1,x_2,\ldots ,x_n$ the program
provides, by using the function \verb SUBlatSUB  of \cite{KATS2},
two $d\times k$ matrices. The rows of the first matrix are the
vectors   \[x_1,x_2,\ldots ,x_n,x_{n+1},\ldots ,x_d,\] while the
rows of the second matrix are the vectors $b_1,b_2,\ldots ,b_d$ of
the positive basis for the subspace $Y$.

Consequently, let $x_1,x_2,...,x_n$ be a collection of linearly
independent, positive vectors  of $\mathbb{R}^k$, then we construct
a matrix $B$ whose columns are the vectors of the given collection
and then we apply the functions \verb SUBlat  and  \verb MINlat  on
that matrix as follows,
\begin{verbatim}
[VectorSublattice,Positivebasis]=SUBlat(B)
[Minimallatticesubspace,Positivebasis]=MINlat(B)
\end{verbatim}

In order to illustrate the most important features of
 \verb SUBlat  and  \verb MINlat  , we reproduce two examples featured in \cite{KATS,POLY9}.

\begin{example}\label{Ex3}  {\em Consider the following  10 vectors $x_1,x_2,...,x_{10}$ in
$\mathbb{R}^{17},$  \[\left[
                     \begin{array}{c}
                        x_1 \\
  x_2 \\
  x_3 \\
  x_4 \\
  x_5 \\
  x_6 \\
  x_7 \\
  x_8 \\
  x_9 \\
  x_{10} \\\end{array}\right]=\left[
  \begin{array}{ccccccccccccccccc}
1 &1& 0& 0& 1& 0& 3& 4& 1& 10& 11& 60& 0& 12& 4 &32 &13 \\ 2 &1 &40&
30& 2& 23& 4& 5& 2& 9& 12& 1& 1& 1& 5& 33& 14\\ 3& 10& 20& 10& 3&
24& 5& 6& 3& 8& 13& 2& 2& 2& 6& 34& 15 \\4& 30& 30& 0& 4& 25& 6& 7&
0& 7& 14& 3& 3& 3& 7& 35& 16 \\5& 40& 1& 0& 5& 0& 7& 8& 5& 6& 15& 0&
0& 3& 8& 0& 17 \\6& 50& 0& 0& 6& 27& 8& 9& 0& 5& 16& 50& 11& 12& 5&
37& 18 \\7& 1& 1& 1& 7& 28& 9& 10& 0& 4& 17& 0& 40& 5& 4& 38& 19\\8&
3& 0& 0& 1& 29& 10& 11& 1& 3& 18& 10& 0& 10& 3& 39& 20 \\ 9& 50& 0&
40& 2& 30& 11& 12& 12& 2& 19& 0& 10& 10& 2& 40& 21 \\ 10& 70& 40& 1&
3& 31& 12& 13& 13& 1& 20& 2& 2& 2& 1& 41& 22\\
  \end{array}
\right]\]

In \cite{KATS} it is provided that the subspace $X=[x_1,x_2,\ldots
,x_{10}]$ is not a lattice-subspace of $\mathbb{R}^{17}.$ In order
to calculate the vector sublattice that $X$ generates as well as a
minimal lattice-subspace containing $X$ we  invoke the
   \verb SUBlat  and the \verb MINlat  functions by typing in the command
window of the Matlab environment:
\begin{verbatim}
>>[VectorSublattice,Positivebasis]=SUBlat(B)
>>[Minimallatticesubspace,Positivebasis]=MINlat(B)
\end{verbatim}
where \verb B  denotes a matrix whose columns are the vectors
$x_1,x_2,\ldots ,x_{10}.$ The results, then, are as follows:
\scriptsize{\begin{verbatim}VectorSublattice =
1  1   0   0  1  0   3  4  1  10  11  60  0  12  4  32  13
2  1   40  30 2  23  4  5  2  9   12  1   1  1   5  33  14
3  10  20  10 3  24  5  6  3  8   13  2   2  2   6  34  15
4  30  30  0  4  25  6  7  0  7   14  3   3  3   7  35  16
5  40  1   0  5  0   7  8  5  6   15  0   0  3   8  0   17
6  50  0   0  6  27  8  9  0  5   16  50  11 12  5  37  18
7  1   1   1  7  28  9  10 0  4   17  0   40 5   4  38  19
8  3   0   0  1  29  10 11 1  3   18  10  0  10  3  39  20
9  50  0   40 2  30  11 12 12 2   19  0   10 10  2  40  21
10 70  40  1  3  31  12 13 13 1   20  2   2  2   1  41  22
0  0   0   0  0  0   0  0  0  0   0   0   0  0   0  0   175
0  0   0   0  0  0   0  0  0  0   0   0   0  0   0  329 0
0  0   0   0  0  0   0  0  0  0   0   0   0  60  0  0   0
0  0   0   0  0  0   0  0  0  0   0   128 0  0   0  0   0
0  0   0   0  0  0   0  0  0  0   155 0   0  0   0  0   0
0  0   0   0  0  0   0  0  0  55  0   0   0  0   0  0   0
0  0   0   0  0  0   0  85 0  0   0   0   0  0   0  0   0

Positivebasis =
0  0   0   0  0  0   0  0   0  0   0   0   69 0   0  0   0
0  0   0   0  0  217 0  0   0  0   0   0   0  0   0  0   0
0  0   132 0  0  0   0  0   0  0   0   0   0  0   0  0   0
0  0   0   82 0  0   0  0   0  0   0   0   0  0   0  0   0
0  256 0   0  0  0   0  0   0  0   0   0   0  0   0  0   0
55 0   0   0  0  0   0  0   0  0   0   0   0  0   0  0   0
0  0   0   0  0  0   0  170 0  0   0   0   0  0   0  0   0
0  0   0   0  0  0   0  0   37 0   0   0   0  0   0  0   0
0  0   0   0  34 0   0  0   0  0   0   0   0  0   0  0   0
0  0   0   0  0  0   0  0   0  0   310 0   0  0   0  0   0
0  0   0   0  0  0   0  0   0  0   0   0   0  0   0  0   350
0  0   0   0  0  0   75 0   0  0   0   0   0  0   0  0   0
0  0   0   0  0  0   0  0   0  0   0   0   0  0   0  658 0
0  0   0   0  0  0   0  0   0  0   0   0   0  0   45 0   0
0  0   0   0  0  0   0  0   0  110 0   0   0  0   0  0   0
0  0   0   0  0  0   0  0   0  0   0   0   0  120 0  0   0
0  0   0   0  0  0   0  0   0  0   0   256 0  0   0  0   0


Minimallatticesubspace =
1   1  0  0  1  0   3   4  1  10 11  60  0   12 4  32  13
2   1  40 30 2  23  4   5  2  9  12  1   1   1  5  33  14
3   10 20 10 3  24  5   6  3  8  13  2   2   2  6  34  15
4   30 30 0  4  25  6   7  0  7  14  3   3   3  7  35  16
5   40 1  0  5  0   7   8  5  6  15  0   0   3  8  0   17
6   50 0  0  6  27  8   9  0  5  16  50  11  12 5  37  18
7   1  1  1  7  28  9   10 0  4  17  0   40  5  4  38  19
8   3  0  0  1  29  10  11 1  3  18  10  0   10 3  39  20
9   50 0  40 2  30  11  12 12 2  19  0   10  10 2  40  21
10  70 40 1  3  31  12  13 13 1  20  2   2   2  1  41  22
0   0  0  0  0  0   10  15 0  55 50  0   0   0  0  0   60
0   0  0  0  0  0   0   0  0  0  0   0   0   60 0  0   0
0   0  0  0  0  0   0   0  0  0  0   128 0   0  0  0   0

Positivebasis =
0  0   0   0  0  0   0  0  0  0   0   0   69 0   0  0   0
0  0   0   0  0  217 0  0  0  0   0   0   0  0   0  0   0
0  0   132 0  0  0   0  0  0  0   0   0   0  0   0  0   0
0  0   0   82 0  0   0  0  0  0   0   0   0  0   0  0   0
0  256 0   0  0  0   0  0  0  0   0   0   0  0   0  0   0
55 0   0   0  0  0   65 70 0  0   105 0   0  0   0  0   115
0  0   0   0  0  0   0  0  37 0   0   0   0  0   0  0   0
0  0   0   0  34 0   0  0  0  0   0   0   0  0   0  0   0
0  0   0   0  0  0   0  0  0  0   0   0   0  0   45 0   0
0  0   0   0  0  0   20 30 0  110 100 0   0  0   0  0   120
0  0   0   0  0  0   0  0  0  0   0   0   0  0   0  329 0
0  0   0   0  0  0   0  0  0  0   0   0   0  120 0  0   0
0  0   0   0  0  0   0  0  0  0   0   256 0  0   0  0   0
\end{verbatim}}}
\end{example}

\begin{example}\label{Ex4}  {\em Consider the following  4 vectors $x_1,x_2,x_3,x_4$
in $\mathbb{R}^{7},$  \[\left[
                     \begin{array}{c}
                        x_1 \\
  x_2 \\
  x_3 \\
  x_4 \\
  \end{array}\right]=\left[
  \begin{array}{cccccccccc}
     1  &   2   &  1&   0 &   1&   1&  4\\
     0  &  1  &   1 &    1&    1 & 0&   2\\
     2  & 1 & 0  &  1    & 1 &   1    & 2  \\
     1  & 0& 1   &  1   &  1   &  0  &  0 \\
  \end{array}
\right]\]

where following the same procedure, as before, one gets

\scriptsize{\begin{verbatim}
VectorSublattice =
1     2     1     0     1     1     4
0     1     1     1     1     0     2
2     1     0     1     1     1     2
1     0     1     1     1     0     0
0     0     0     0     0     2     0
0     4     0     0     0     0     8

Positivebasis =
0     0     0     3     0     0     0
0     0     0     0     0     4     0
4     0     0     0     0     0     0
0     8     0     0     0     0     16
0     0     0     0     4     0     0
0     0     3     0     0     0     0

Minimallatticesubspace =
1    2    1    0    1    1    4
0    1    1    1    1    0    2
2    1    0    1    1    1    2
1    0    1    1    1    0    0
0    4    0    0    0    0    8

Positivebasis =
0    0    0    3    1.5  0    0
0    8    0    0    0    0    16
4    0    0    0    0    0    0
0    0    3    0    1.5  0    0
0    0    0    0    1    2    0
\end{verbatim}}}
\end{example}

\begin{remark}
\em{ In the beginning of the present section, we mentioned that in
order to determine the vectors $\xi_i:D(\beta)\rightarrow
\mathbb{R}_+, i=1,2,\ldots ,d$ the function \verb MINlat  responds
by solving $k$ underdetermined $n\times d$ linear systems subject to
the inequalities $\xi_i\geq 0$ ,$i=1,\ldots ,n.$  Since, the
solution to such systems is not unique then, if $p$ denotes a
solution corresponding to one of these systems, then $y=p+h$, where
$h$ is an arbitrary vector from the null space, is a solution too.
Thus, a minimal lattice-subspace containing the vectors
$x_1,x_2,...,x_n$ is not unique. In \cite{POLY9}, Example 3.21, it
is proved that for the collection of vectors of Example \ref{Ex4}
there exist two minimal lattice-subspaces $Y,Y'$ such that
\begin{itemize}
\item $Y\neq Y'$,
\item the space $Y\cap Y'$ is not a lattice-subspace,
\item and $Y,Y'$ are not subspaces of $Z$ ($Z$ denotes the generated vector sublattice).
\end{itemize}
}
\end{remark}

\subsection{Comparison Results}\label{SS1}
For the purpose of monitoring the performance, in this section we
present  the execution times of the proposed methods
 (\verb SUBlat , \verb MINlat  ) for various collections of vectors
and dimensions. For this purpose we have used the function
    \verb testMINlat  , \verb testSUBlat  (see Appendix) in order to test $50$ full rank matrices
for each rank $n$, $n=3,...,30$. The cumulative results are
presented in Table 1 while, the time responses have been recorded
using the Matlab function {\bf profile}.

It is evident, from Table 1, that the proposed numerical methods,
based on the introduction of the
 \verb SUBlat  and the \verb MINlat  function, enable us to perform fast
estimations for a variety of dimensions.

\begin{table}[h]
\label{v1} \caption{Results for 50 tested full rank matrices for
each rank $n$, $n=3,...,30.$ } \scriptsize{\begin{tabular}{cccccc}
  \hline
  Rank & \verb SUBlat  & \verb MINlat   & Rank &\verb SUBlat &\verb MINlat  \\
 & (Total time in seconds) & (Total time in seconds) & & (Total time in
seconds) & (Total time in seconds)\\
  \hline
  3 &0.553&0.689&17&2.812&1.528\\
  4&0.569&0.705&18&2.994&1.732\\
  5&0.584&0.742&19&3.333&1.885\\
  6&0.696&0.822&20&3.594&2.222\\
  7&0.856&0.831&21&4.084&2.428\\
  8&0.973&0.890&22&4.219&2.773\\
  9&1.141&0.903&23&4.569&3.175\\
  10&1.307&1.012&24&4.817&3.695\\
  11&1.642&1.047&25&5.149&4.306\\
  12&1.667&1.108&26&5.712&4.815\\
  13&1.890&1.151&27&5.910&5.741\\
  14&2.059&1.185&28&6.229&6.566\\
  15&2.314&1.270&29&6.675&7.493\\
  16&2.765&1.365&30&7.003&9.226\\
 \hline
\end{tabular}}
\end{table}

\section{Applications in economies with incomplete asset markets}\label{S3}
 The theory of vector sublattices and lattice-subspaces
has been extensively used in the last years in Mathematical
Economics, especially in the areas of incomplete markets and
portfolio insurance. Recently, in \cite{KOUN}, a new approach to the
problem of completion by options of a two-period security market was
introduced, which used the theory of positive bases in vector
sublattices of $\mathbb{R}^k$ (as described in \cite{POLY9}). Also,
if $x_1,x_2,\ldots ,x_n$ denotes a collection of linearly
independent, positive vectors, of $\mathbb{R}^k$ then if
$X=[x_1,x_2,\ldots ,x_n]$ is a lattice subspace or, in the contrary
case, if there exists a minimal lattice-subspace containing $X$ then
a solution to  a cost minimization problem known as minimum-cost
portfolio insurance always exists. In this section we shall discuss
these interconnections and briefly describe the theoretical
background on this subject. A full theoretical development is
available in
\cite{ALIP4,ALIP2,ALIP3,KATS,KATS1,KOUN,POLY1,POLY8,POLY9}. We will
be also presenting how one can use the proposed functions,
 \verb SUBlat  and \verb MINlat , in order to determine the completion of security markets as well as  the solution of
 the minimum-cost portfolio insurance problem.

\subsection{Completion  of security markets}
The problem of completion by options of a two-period security market
in which the space of marketed securities is a subspace of
$\mathbb{R}^k$ has been studied in \cite{KOUN}. The present study
involves vector sublattices generated by a subset $B$ of
$\mathbb{R}^k$ of positive, linearly independent vectors, so we
shall provide a computational solution to this problem by using the
\verb SUBlat function  in order to provide the generated vector
sublattice $Y$ as well as a positive basis for $Y.$

Let us assume that in the beginning of a time period there are $n$
securities traded in a market. Let $\mathcal{S}=\{1,...,k\}$ denote
a finite set of states and $x_j\in \mathbb{R}_+^k$ be the payoff
vector of security $j$ in $k$ states. The payoffs $x_1,x_2,...,x_n$
are assumed linearly independent so that there are no redundant
securities. If $\theta=(\theta_1,\theta_2,...,\theta_n)\in
\mathbb{R}^n$ is a non-zero portfolio then its payoff is the vector
$$T(\theta)=\sum_{i=1}^n\theta _i x_i.$$
The set of payoffs of all portfolios is referred as the space of
\textit{marketed securities} and it is the linear span of the
payoffs vectors $x_1,x_2,...,x_n$ in $\mathbb{R}^k$ which we shall
denote it by $X$, i.e.,
\[X=[x_1,x_2,...x_n].\]  For any $x,u \in \mathbb{R}^k$
and any real number $a$ the vector $c_u(x,a)=(x-au)^+$ is the
\textit{call option} and  $p_u(x,a)=(au-x)^+$ is the \textit{put
option} of $x$ with respect to the \textit{strike vector} $u$ and
\textit{exercise price} $a$.

Following the terminology of \cite{KOUN}, let $U$ be a fixed
subspace of $\mathbb{R}^k$ which  is called \textit{strike subspace}
and the elements of $U$ are the \textit{strike vectors}. Then, the
\textit{completion by options} of the subspace $X$ with respect to
$U$ is the space $F_U(X)$ which is defined inductively as follows:
\begin{itemize}
  \item $X_1$ is the subspace of $\mathbb{R}^k$ generated by $\mathcal{O}_1,$ where $\mathcal{O}_1=\{c_u(x,a)|x\in X,u\in U,a\in \mathbb{R}\}$,
  denotes the set of call options written on the elements of $X$,
  \item $X_n$ is the subspace of $\mathbb{R}^k$ generated by $\mathcal{O}_n,$ where $\mathcal{O}_n=\{c_u(x,a)|x\in X_{n-1},u\in U,a\in \mathbb{R}\}$,
  denotes the set of call options written on the elements of $X_{n-1}$,
  \item $F_U(X)=\cup _{n=1}^{\infty} X_n$.
  \end{itemize}

The completion by options $F_U(X)$ of $X$ with respect to $U$  is
the vector sublattice of $\mathbb{R}^k$ generated by the subspace
$Y=X\cup U$. The details are presented in the next theorem,

\begin{theorem}\cite[\textit{Theorem 3}]{KOUN}\label{T4}
In the above notation, we have
\begin{enumerate}
  \item [$(i)$] $Y\subseteq X_1$,
  \item [$(ii)$] $F_U(X)$ is the sublattice $S(Y)$ of  $ \mathbb{R}^k$
  generated by $Y$, and
  \item [$(iii)$] if $U\subseteq X$, then $F_U(X)$ is the sublattice
  of $\mathbb{R}^k$ generated by $X.$
\end{enumerate}
\end{theorem}

\begin{definition}\cite[\textit{Definition 10}]{KOUN}\label{D1}
Any set $\{y_1,y_2,\ldots ,y_r\}$ of linearly independent positive
vectors of $\mathbb{R}^k$ such that $F_U(X)$ is the sublattice of
$\mathbb{R}^k$ generated by $\{y_1,y_2,\ldots ,y_r\}$ is a {\bf
basic set } of the market.
\end{definition}

\begin{theorem}\cite[\textit{Theorem 11}]{KOUN}\label{T6} Any
maximal subset $\{y_1,y_2,\ldots ,y_r\}$ of linearly independent
vectors of $\mathcal{A}$  is a basic set of the market, where
$\mathcal{A}=\{x_1^+,x_1^-,\ldots , x_n^+,x_n^-\}$, if  $U\subseteq
X$ and $\mathcal{A}=\{x_1^+,x_1^-,\ldots ,
x_n^+,x_n^-,u_1^+,u_1^-,\ldots ,u_d^+,u_d^-\}$, if $U\subsetneq X$

\end{theorem}

\begin{definition}\cite[\textit{Definition 12}]{KOUN}\label{D2}
The space of marketed securities  $X$ is {\bf complete by options}
with respect to $U$ if $X=F_U(X)$.
\end{definition}

From theorem \ref{T3} and definition \ref{D2} it follows,

\begin{theorem}\cite[\textit{Theorem 13}]{KOUN}\label{T5} The space
$X$ of marketed securities is complete by options with respect to
$U$ if and only if $U\subseteq X$ and $cardR(\beta )=n$.
\end{theorem}

\begin{theorem}\cite[\textit{Theorem 14}]{KOUN}\label{T6} The dimension of $F_U(X)$ is equal to the cardinal number of $R(\beta
)$. Therefore, $F_U(X)=\mathbb{R}^k$ if and only if
$cardR(\beta)=k.$
\end{theorem}

In order to apply our method to the problem of completion of
security markets we present the following example which is formerly
featured in \cite{KOUN}.

\begin{example}\cite[\textit{Example 16}]{KOUN}\label{Ex5}
\em{Suppose that in a security market, the payoff space is
$\mathbb{R}^{12}$ and the primitive securities are:

\[x_1=(1,2,2,-1,1,-2,-1,-3,0,0,0,0)\]
\[x_2=(0,2,0,0,1,2,0,3,-1,-1,-1,-2)\]
\[x_3=(1,2,2,0,1,0,0,0,-1,-1,-1,-2)\]
and that the strike subspace is the vector subspace $U$ generated
 by the vector \[u=(1,2,2,1,1,2,1,3,-1,-1,-1,-2).\] Then, a maximal
 subset of linearly independent vectors of
 $\{x_1^+,x_1^-,x_2^+,x_2^-,x_3^+,x_3^-,u_1^+,u_1^-\}$ can be
 calculated by using the following code\footnote{The rref function is a
 Matlab function that produces the reduced
row echelon form of a given matrix by using Gauss Jordan elimination
with partial pivoting (cf. \cite{MATL1}). }:

\begin{verbatim}
>>XX = [max(X,zeros(size(X)));max(-X,zeros(size(X)))];
>>S = rref(XX');
>>[I,J] = find(S);
>>Linearindep = accumarray(I,J,[rank(XX),1],@min)';
>>W = XX(Linearindep,:);
\end{verbatim}
where $X$ denotes a matrix whose rows are the vectors
$x_1,x_2,x_3,u.$ We can determine the completion by options of $X$
i.e., the space $F_U(X)$, with the  \verb SUBlat  function by using
the following code:
\begin{verbatim}
>>[VectorSublattice,Positivebasis]=SUBlat(W')
\end{verbatim}
The results then are as follows

\scriptsize{\begin{verbatim}
VectorSublattice =
1     2     2     0     1     0     0     0     0     0     0     0
0     2     0     0     1     2     0     3     0     0     0     0
1     2     2     1     1     2     1     3     0     0     0     0
0     0     0     0     0     0     0     0     1     1     1     2
2     0     4     0     0     0     0     0     0     0     0     0

Positivebasis =
0     0     0     0     0     0     0     0     1     1     1     2
0     0     0     1     0     0     1     0     0     0     0     0
0     0     0     0     0     4     0     6     0     0     0     0
4     0     8     0     0     0     0     0     0     0     0     0
0     6     0     0     3     0     0     0     0     0     0     0
\end{verbatim}}

 }
\end{example}

\subsection{Minimum cost portfolio insurance}
In this section we shall, briefly discuss an investment strategy
called minimum-cost portfolio insurance as a solution of a cost
minimization problem. We will be also presenting how  one can use
the previous results for the calculation of the minimal
lattice-subspace, in order to calculate the minimum-cost insured
portfolio. In our model we use a method of comparing portfolios
called portfolio dominance ordering. This ordering  compares
portfolios by means of the ordering of their payoffs. Under this
consideration we are able to use the order structure of the payoff
space together with the theory of lattice-subspaces. In what follows
we shall use the notation introduced in \cite{POLY1}.

The model of security markets we study here is extended over two
periods, namely  period $0$ and period $1.$ We
 assume  $n$ securities labeled by the natural numbers
$1,2,...,n$, acquired during the period $0$ and that these $n$
securities are described by their payoffs at date $1.$ The payoff of
the $ith$ security is in general a positive element $x_i$ of an
ordered space $E$ which is called \textit{payoff space}. In
addition, we assume that the payoffs $x_1,x_2,...,x_n$ are linearly
independent so that there are no redundant securities and that the
securities have limited liability which ensures the positivity of
$x_1,x_2,...,x_n.$ In \cite{POLY1}, it is assumed that $E$ is the
space of real valued continuous functions $C(\Omega)$ defined in a
compact, Hausdorff topological space $\Omega$. A portfolio is a
vector $\theta=(\theta _1,\theta _2,...,\theta _n)$ of
$\mathbb{R}^n$ where $\theta _i$ is the number of shares of the
$ith$ security. The space $\mathbb{R}^n$ is then known as
\textit{portfolio space}. If
$\theta=(\theta_1,\theta_2,...,\theta_n)\in \mathbb{R}^n$ is a
non-zero portfolio then its payoff is the vector
$$R(\theta)=\sum_{i=1}^n\theta _i x_i\in C(\Omega).$$
The operator $R$ is one-to-one and is called the \textit{payoff
operator}. The pointwise ordering in $C(\Omega),$ induces the
partial ordering $\geq _R$ in the portfolio space $\mathbb{R}^n$ and
is defined as follows: For each $\theta,\phi\in \mathbb{R}^n$ we
have
$$\theta \geq _R \phi, \textit{   if and only if  } R(\theta)\geq
R(\phi).$$ This ordering is known as the \textit{portfolio dominance
ordering}. The set of payoffs of all portfolios, or the range space
of the payoff operator, is the linear span of the payoffs vectors
$x_1,x_2,...,x_n$ in $C(\Omega)$ which we shall denote it by
$\mathcal{M}$, i.e.,
$$\mathcal{M}=[x_1,x_2,...,x_n].$$ The subspace $\mathcal{M}$ of $C(\Omega)$
is called the \textit{asset span} of securities or the \textit{space
of marketed securities}.

Let us assume that $p=(p_1,p_2,...,p_n)\in \mathbb{R}^k$ is a vector
of security prices and $\theta,\phi$ are two portfolios. Then, the
insured payoff on the portfolio
$\theta=(\theta_1,\theta_2,...,\theta_n)$ at the ''floor'' $\phi$
and in the price $p$ is the contingent claim $R(\theta)\vee
R(\phi).$

The solution of the following cost minimization problem is referred
to as the \textit{minimum-cost insured portfolio}, or a
\textit{minimum-cost insurance of the portfolio $\theta$ at the
floor $\phi$ and in the price $p,$}
$$\min_{\eta \in \mathbb{R}^k} p\cdot \eta$$ subject to
$$R(\eta)\geq R(\theta)\vee R(\phi).$$

In \cite{POLY1} it is proved that if the payoff space is contained
in a minimal lattice-subspace of $C(\Omega)$ then a minimum-cost
insurance of the portfolio $\theta$ always exists. The details are
included in the next theorem:

\begin{theorem}\cite[\textit{Theorem 15}]{POLY1}\label{T7}
If the payoff space $X$ is contained in a finite-dimensional minimal
lattice-subspace $Y$ of $C(\Omega)$ and the sum of the payoff
vectors $x_i$ is strictly positive, then a minimum-cost insurance of
the portfolio $\theta$ at the floor $\phi$ and in the price $p,$
exists and it is determined by solving the corresponding
minimization problem.
\end{theorem}
 In order to apply our method,
 \verb MINlat , we consider that $\Omega =\{1,2,...,n\}$. Then, it
 is evident that $C(\Omega)=\mathbb{R}^n.$ Therefore,  in view of Theorem \ref{T7} and by using the \verb MINlat  function
 we are able to determine the minimum-cost insurance of the
 given portfolio $\theta$ at the
floor $\phi$ and in the price $p.$

\section{Conclusions}
In this paper, new computational methods in order to determine
vector sublattices and minimal lattice-subspaces of $\mathbb{R}^k$
are presented. In order to reach our goal the study of a
vector-valued function $\beta$ is further involved by introducing
two Matlab functions, namely \verb SUBlat , and \verb MINlat . The
results of this work can give us an important tool in order to study
the interesting problems of completion by options of a two-period
security market in which the space of marketed securities is a
subspace of $\mathbb{R}^k$ and in portfolio insurance. The
experiment results, in subsection \ref{SS1}, show that our algorithm
performs well. Also, note that the algorithm \ref{S13} determines
the exact solution to the problems of completion by options and the
minimum-cost insured portfolio. Finally we are convinced that, from
the mathematical point of view, the proposed algorithm can be
further analyzed independently, in terms of formal numerical
analysis.
\section{Appendix}
The \verb SUBlat  function

\begin{verbatim}
function [Sublattice,Positivebasis] = SUBlat(B)
%SUBlat(B) provides the vector sublattice generated by
%a given finite collection of positive, linearly
%independent vectors of R^n
%B denotes the matrix whose columns are the given vectors
[N,M] = size(B);
Id = eye(N);
for i = 1:N,
  if norm(B(i,:),1)~=0,
Test(i,:) = 1/norm(B(i,:),1)*B(i,:);
  end
end
Matrix = Test;
[BB,m,n] = unique(Matrix,'rows');
Index = 1:N;
S = rref(BB');
[I,J] = find(S);
Linearindep = accumarray(I,J,[rank(BB),1],@min)';
mm = length(m);
nn = length(n);
Index1 = 1 : mm;
Index2 = setdiff(Index1,Linearindep);
YY = sum(B,2)';
TTT = setdiff(Index,m(Linearindep));
KK = Id(TTT,:);
TT = YY(1,TTT)';
T = diag(TT)*KK;
K = zeros(N);
K(TTT,:) = T;
Vec = zeros(mm-M,N);
if mm < nn,
 for i = 1:length(Index2),
     DD = strmatch(Index2(i),n,'exact')' ;
     R = length(DD);
    if R >= 2,
       Vector = sum(K(DD,:));
    else
       Vector = K(DD,:);
    end
       Vec(i,:) = Vector;
 end
      [a,b] = find(Vec);
      Vectors = Vec(unique(a),:);
      Sublattice = [B';Vectors];
      Positivebasis = SUBlatSUB(Sublattice');
else
      KKK = unique(K,'rows');
      [II,JJ] = find(KKK);
      Vectors = KKK(unique(II),:);
      Sublattice = [B';Vectors];
      Positivebasis = SUBlatSUB(Sublattice');
end
\end{verbatim}

The \verb MINlat  function

\begin{verbatim}
function [MLS,Positivebasis] = MINlat(B)
%MINlat(B) provides the minimal lattice-subspace
%generated by a given finite collection of positive,
%linearly independent vectors of R^n
%B denotes the matrix whose columns are the given vectors
[N,M] = size(B);
for i=1:N,
  if norm(B(i,:),1)~=0,
    Test(i,:) = 1/norm(B(i,:),1)*B(i,:);
  end
end
Matrix = Test;
[ii,jj] = find(Matrix);
Matrix1 = Matrix(unique(ii),:);
BB = unique(Matrix1,'rows');
M1 = rank(bsxfun(@minus,BB,BB(1,:)));
if M1<M,
  Utrans = bsxfun(@minus,BB,BB(1,:));
  Rot = orth(Utrans');
  Uproj = Utrans*Rot;
  Tri = convhulln(Uproj);
  VIndex = unique(Tri(:));
  P = BB(VIndex,:)';
  Q = length(VIndex);
else
  VIndex = unique(convhulln(BB));
  P = BB(VIndex,:)';
  Q = length(VIndex);
end
  Test = zeros(N,M);
for i=1:N,
  Test(i,:) = 1/norm(B(i,:),1)*B(i,:);
end
  Matrix = Test;
  BBB=Matrix';
  R = Q-M;
  Sol = zeros(Q,R);
for i = 1:N,
  Sol(:,i) = lsqnonneg(P,BBB(:,i));
end
  Solutions = Sol;
  Norms = sum(B,2)';
  Test1 = zeros(R,N);
for i = 1:R,
  Index1 = M+i;
  D = Solutions(Index1,:).*Norms*eye(N);
  Test1(i,:) = D;
end
  Minlatsub = [B';Test1];
  MLS = Minlatsub;
  Positivebasis = SUBlatSUB(MLS');
\end{verbatim}

The \verb SUBlatSUB  function

\begin{verbatim}
function [positivebasis,dimensions] = SUBlatSUB(A)
%SUBlatSUB(A)  provides the vector sublattice
%or the lattice-subspace of a given finite collection
%of positive, linearly independent vectors of R^n
%A denotes the matrix whose columns are the given vectors
if any(any(A<0))~=0,
    error('the initial matrix must have positive elements')
end
[N,M] = size(A);
if rank(A)~=M,
    error('the given vectors are linearly dependent')
end
for i=1:N,
 if norm(A(i,:),1)~=0,
  Test(i,:) = 1/norm(A(i,:),1)*A(i,:);
 end
end
  matrix = Test;
 [ii,jj] = find(matrix);
  matrix1 = matrix(unique(ii),:);
  u = unique(matrix1,'rows');
  m = length(u(:,1));
if M == m,
  disp('vector sublattice')
  positivebasis = inv(u')*A';
  dimensions = [M m N]';
else
  m1 = rank(bsxfun(@minus,u,u(1,:)));
  if  m1<M,
   utrans = bsxfun(@minus,u,u(1,:));
   rot = orth(utrans');
   uproj = utrans*rot;
   tri = convhulln(uproj);
   d = length(unique(tri(:)));
   if d == M,
   basis = inv(u(unique(tri(:)),:)')*A';
   disp('lattice-subspace')
   positivebasis = basis;
   dimensions = [M m d N]';
   else
    disp('not a lattice-subspace')
   dimensions = [M m d N]';
   positivebasis=[];
   end
  end
end
\end{verbatim}

The \verb testSUBlat  function

\begin{verbatim}
function testSUBlat = testSUBlat(k,j)
%k is the the dimension of the Euclidean space R^k
%j is the number of the tested matrices
for i = 1:j,
    A = rand(k+2,k);
    testSUBlat = SUBlat(A);
end
\end{verbatim}

The \verb testMINlat  function

\begin{verbatim}
function testMINlat = testMINlat(k,j)
%k is the the dimension of the Euclidean space R^k
%j is the number of the tested matrices
for i = 1:j,
    A = rand(k+2,k);
    testMINlat = MINlat(A);
end
\end{verbatim}

{\bf Acknowledgments}\\ The research of the first author was
financially supported by the State Scholarship Foundation (IKY) in
his postdoctoral studies.



\begin{thebibliography}{00}
\bibitem{ABRA2}
     Y.A. Abramovich, C.D. Aliprantis and I.A. Polyrakis, Lattice-Subspaces and positive
     projections, {\em Proc.R.Ir.Acad., {\bf 94A}}(1994), 237-253.

\bibitem{ALIP4}
     C.D. Aliprantis, D.J. Brown and J. Werner, Incomplete derivative markets and portfolio insurance,
      {\em Cowles Foundation Discussion Paper, {\bf 1126R}}(1997), 1-13.

\bibitem{ALIP2}
     C.D. Aliprantis, D.J. Brown and J. Werner, Minimum-cost portfolio
     insurance,
      {\em Journal of Economic Dynamics \& Control, {\bf 24}}(2000), 1703-1719.

\bibitem{ALIP3}
     C.D. Aliprantis, I.A. Polyrakis and R. Tourky, The cheapest
     hedge, {\em Journal of Mathematical Economics, {\bf 37}}(2002), 269-295.


\bibitem{MATL1}
MATLAB v 7.4(R2007b), Help Browser.

\bibitem{MATL2}
MATLAB User's Guide, The MathWorks Inc.

\bibitem{KATS} V.N. Katsikis, Computational methods in portfolio insurance, {\em Applied Mathematics and
Computation, {\bf 189}}(2007), 9-22.

\bibitem{KATS1} V.N. Katsikis, Computational methods in lattice-subspaces of $C[a,b]$ with
applications in portfolio insurance, {\em Applied Mathematics and
Computation, {\bf 200}}(2008), 204-219.

\bibitem{KATS2} V.N. Katsikis, A Matlab-based rapid method for computing lattice-subspaces and vector
sublattices of $\mathbb{R}^n$: Applications in portfolio insurance,
{\em Applied Mathematics and Computation, {\bf 215}}(2009) 961–972.

\bibitem{KOUN} C. Kountzakis, I.A. Polyrakis, The completion of security
markets, {\em Decisions in Economics and Finance, {\bf
29}}(2006),1-21.

\bibitem{POLY1} I.A. Polyrakis, Linear Optimization in $C(\Omega)$ and Portfolio Insurance,
{\em Optimization {\bf 52}}(2003), 221-239.

\bibitem{POLY8} I.A. Polyrakis, Finite-dimensional
lattice-subspaces of $C(\Omega)$ and curves of $\mathbb{R}^{n}$,
{\em Transactions of the American Mathematical Society {\bf
348}}(1996), 2793-2810.

\bibitem{POLY9} I.A. Polyrakis, Minimal lattice-subspaces,
{\em Transactions of the American Mathematical Society {\bf
351}}(1999), 4183-4203.

\end{thebibliography}
\end{document}